\documentclass[12pt]{article}

\usepackage{amsfonts}
\usepackage{bbm}
\usepackage{cite}

\newcommand{\bmat}{\left(\begin{array}}
\newcommand{\emat}{\end{array}\right)}

\def\gtrsim{\mathrel{\raise.3ex\hbox{$>$\kern-.75em\lower1ex\hbox{$\sim$}}}}
\def\NPB#1#2#3{Nucl. Phys. B{#1} (#2) #3}

\def\-{\hphantom{-}}

\def\s2{\frac{1}{\sqrt2}}

\def\beq{\begin{equation}}
\def\eeq{\end{equation}}
\def\beqa{\begin{eqnarray}}
\def\eeqa{\end{eqnarray}}

\def\mg{m_{3/2}}
\def\mg2{m^2_{3/2}}

\def\Dsl{\,\raise.15ex\hbox{/}\mkern-13.5mu D} 

\begin{document}
\pagestyle{plain}

\makeatletter
\makeatother
\pagestyle{empty}
\rightline{ IFT-UAM/CSIC-06-06}
\begin{center}
\LARGE{ Flux-Induced Baryon Asymmetry 
\\[10mm]}
\large{ L.~E.~Ib\'a\~nez  \\[6mm]}
        \small{ Departamento de F\'{\i}sica Te\'orica C-XI
        and \\ Instituto de F\'{\i}sica Te\'orica  C-XVI\\
        Universidad Aut\'onoma de Madrid\\
        Cantoblanco, 28049 Madrid, Spain.\\[8mm]}
\small{\bf Abstract} \\[8mm]
\end{center}
{\small
I propose that the primordial baryon asymmetry 
of the universe was induced by the
presence of a non-vanishing antisymmetric field background $H_{\mu \nu \rho}$
across the three space dimensions. 
This background creates a dilute $(B-L)$-number density in the universe  
 cancelling  the contribution from baryons and leptons.
 This situation naturally appears if the   $U(1)_{B-L}$
symmetry is gauged and the corresponding gauge boson gets
a Stuckelberg mass by combining with an antisymmetric
field $B_{\mu \nu }$. 
All these ingredients are present in  D-brane
 models of particle physics. None of the Sakharov conditions
are required.
}

\vspace{5.0cm}
\rightline{\it Dedicado a Leman}


\newpage
\setcounter{page}{1}
\pagestyle{plain}
\renewcommand{\thefootnote}{\arabic{footnote}}


One of the most pressing cosmological puzzles is the
observed baryon asymmetry in the universe. The traditional
solution to this puzzle goes through the triplet of
Sakharov conditions: a baryon asymmetry may be dynamically
generated if the three ingredients 1) Baryon number violation,
2) C and CP-violation and 3) departure from thermal equilibrium
take place  in the history of the universe. A variety of
concrete models which realize this general recipe have
been proposed in the last 30 years.
Although this recipe seems to work, one has the feeling that
the way the baryon asymmetry appears in the history of the
universe in this scheme depends very much on details of the models
and is  certainly   not generic in particle physics
models.

The philosophy underlying the idea of baryogenesis is that
the primordial universe had the quantum numbers of the vacuum and hence
it is natural to assume an exactly  vanishing primordial baryon number
and expect equal numbers of baryons and antibaryons.
Assuming a small missmatch   
of order $10^{-10}$ for $n_B/n_\gamma $
as an initial condition is then totally unnatural.

The purpose of this note  is to point out that
the primordial universe could have the quantum numbers of the
vacuum and still posses a primordial baryon-antibaryon asymmetry.
The idea is that there could be a diluted  distribution
of baryon number density in the vacuum precisely cancelling 
the baryon number of baryons themselves
\footnote{Perhaps an appropriate name for this could be
` Baryonic Aether'.}. Specifically,
I point out that under certain circumstances
a constant
non-vanishing antisymmetric field background $H_{\mu \nu \rho}$
across the three space dimensions may be such an extra source of
baryon (or rather B-L in the example discussed )  number. In this scheme the
overall primordial 
B-L number
vanishes. However, if a $H_{\mu \nu \rho}$ background is present,
a net non-vanishing  $B-L$  from baryons/leptons  
must be also present, due to $U(1)_{B-L}$ conservation.
From this point of view having a B-L 
asymmetry as an initial condition is something generic but
still compatible with vanishing quantum numbers for
the primordial universe. At lower temperatures 
electroweak instanton effects (violating the combination  B+L) will
force to have $n_B=-n_L$,  but will be unable to erase the
baryon and lepton asymmetries.
Notice that  the   Sakharov conditions are not needed.

The essential idea is inspired by the generic phenomenon
in D-brane string compactifications 
(for reviews and references see e.g.
\cite{blumenhagen,kiritsis,uranga,marchesano})
by which  $U(1)$ 
D-brane gauge bosons get Stuckelberg masses by combining with 
antisymmetric $B_{\mu \nu }$ fields. In this case the 
$U(1)$ symmetry survives as a global symmetry below
the scale of the gauge boson mass. We here consider the case 
in which  in addition there is a constant flux
$H_{\mu \nu \rho }$ along the three space dimensions.
We find that in this case the  latter gives rise to a density of $U(1)$
charge proportional to the flux.
A natural gauged $U(1)$ symmetry to consider 
within the context of the SM is B-L,
which is global symmetry of the SM.
In the presence of 3 right-handed neutrinos 
this symmetry is anomaly-free and can thus be gauged without
problems.
This is in fact something generic e.g. in D-brane models of
particle physics 
\cite{blumenhagen,kiritsis,uranga,marchesano}.
In such models the QCD gauge
bosons come 
from three coincident parallel D-branes, giving rise to a   
gauge group $U(3)=SU(3)_{QCD}\times U(1)_B$. The $U(1)_B$
 gauge symmetry corresponds to gauged baryon number.
In the same way lepton number is the $U(1)$ living on the
worldvolume of a corresponding leptonic D-brane \cite{imr}.

In fact a gauged $U(1)_{B-L}$ appears also naturally in 
left-right symmetric extensions of the SM as well as $SO(10)$ 
GUT models.
However there is an important difference with 
our approach here. In the latter models the  $U(1)_{B-L}$ symmetry
is assumed to be spontaneously broken by the standard Higgs mechanism. 
 In  here we are going to assume that  the  source 
for the mass of the $U(1)_{B-L}$ gauge boson will be a 
Stuckelberg mass obtained by combining with an antisymmetric tensor
$B_{\mu \nu }$.
Indeed, one interesting  phenomenon observed while 
constructing SM-like intersecting D-brane models \cite{imr} is that 
often the gauge boson associated to the symmetry $U(1)_{B-L}$
becomes massive by combining with an antisymmetric 
$B_{\mu \nu }$ field.
 A coupling of type $B\wedge F$ between the antisymmetric
field $B$ and the Abelian field strength $F$ is the origin of
this effect.
As we review below, this term mixes both
fields  rendering the gauge boson massive
by swallowing the $B$ field.
Note that in general this mechanism by itself does not give
masses to neutrinos. We will assume that eventually some mechanism will give them
either a (B-L)-preserving Dirac mass or else Majorana masses. As 
in the leptogenesis scenarios, for the observed values of neutrino masses,
these Majorana masses are too small   to erase 
the original B-L asymmetry.

It was known since a long time ago \cite{dsw} that 
in string models there are pseudoanomalous $U(1)$'s 
in which the triangle anomalies are cancelled by the exchange of 
some antisymmetric $B_{\mu \nu }$ field  with appropriate couplings
to the gauge bosons. This is the generalized Green-Schwarz mechanism
\cite{gs} present in 
4-dimensional string compactifications
\cite{dsw}. This works through the
existence of two couplings,  a $B\wedge F$ coupling as mentioned above
 and a  coupling of type $\eta F\wedge F$, where $F$ is any gauge
field strength in the theory and $\eta $ 
is the Poincare dual of $B_{\mu \nu}$. 
 Under a $U(1)$ gauge transformation of parameter
$\theta (x)$ it transforms like $\eta \rightarrow \eta +\theta(x)$.
The combination of both terms renders the corresponding  
compactifications anomaly-free.
 In generic string models there
may be a number of these `pseudo-anomalous $U(1)$´s '  which get
anomaly-free making use of several antisymmetric fields $B$
\cite{iru}.  As we said, the first of these terms, the $B\wedge F$
 has the effect of rendering massive the corresponding 
pseudoanomalous $U(1)$. All pseudoanomalous $U(1)$'s in 
D=4 string compactifications become massive in this way. 
But the reverse is not true: an anomaly free 
$U(1)$  gauge boson (like B-L) may become massive if the appropriate
  $B\wedge F$ coupling is present. In ref.\cite{imr}
it was shown that indeed those couplings are present 
in e.g., intersecting D6 brane models of particle 
physics. More recently it  has also been found that in $N=1$ SUSY
models with  MSSM-like spectra obtained from Type II rational 
CFT orientifolds,  $U(1)_{B-L}$ gauge bosons do also get
Stuckelberg masses in this fashion \cite{dhs}.

As I said the would be anomalous $U(1)_{B-L}$ gauge boson gets a
mass of order the string scale $M_s$. On the other hand
the corresponding $U(1)_{B-L}$ symmetry survives as a {\it global}  
rather than local symmetry   in the
low-energy lagrangian \cite{Witten},\cite{iq,imr}. 
This residual global $U(1)$ symmetries are  rather generic in type II string
D-brane models. It is not the case in traditional heterotic compactifications
on CY with $SU(N)$ bundles.  In the latter  case there is a unique
$B_{\mu \nu }$ field with indices in Minkowski which participates
 in the Green-Schwarz mechanism. A non-vanishing Fayet-Iliopoulos term is
then induced \cite{dsw} which forces some of the charged matter scalars
carrying anomalous $U(1)$ charge to get vevs resulting in
a breaking of the would-be global $U(1)$ symmetry.
In the type II orientifold case such FI-terms  vanish, as long as
SUSY-breaking effects induce masses to scalars charged under the $U(1)$. Thus 
the global $U(1)$ symmetry remains perturbatively unbroken
\footnote{In heterotic compactifications with $U(N)$ bundles 
 instead of $SU(N)$ bundles  \cite{bhw}
 one  gets again a situation similar to that in type II D-brane
models with several anomalous $U(1)$'s and axion-like fields.}.

Let us first recall  how   $U(1)$ gauge bosons gets massive
before adding the H-flux.
Consider the action of  a  gauge boson and an
antisymmetric field $B_{\mu \nu }$  above. 
\footnote{In  e.g. intersecting D6-brane models 
\cite{dbrane,kiritsis,imr,blumenhagen,uranga,marchesano}  such 2-forms  
would be linear combinations of forms arising
 from wrapping the RR 5-form over  3-cycles  in the CY.
Then the flux considered would correspond to a RR flux $F_{6}$
with 3 legs on the 3-cycle and the other 3 on the space dimensions.
Thus the $H_{\mu \nu \rho }$ flux here   is not a NS flux.}.
The relevant piece of the Lagrangian has the form 
\begin{equation}
{\cal L}_0\ =\ -\frac{1}{12} H^{\mu\nu\rho}
H_{\mu\nu\rho}-\frac{1}{4g^2} F^{\mu\nu} F_{\mu\nu}
+ \frac{cM}{4}\ \epsilon^{\mu\nu\rho\sigma} B_{\mu\nu}\ F_{\rho\sigma}
\ 
\label{lagrana}
\end{equation}
where 
\begin{equation}
H_{\mu \nu\rho}=\partial_\mu
B_{\nu\rho}+\partial_\rho B_{\mu\nu}+\partial_\nu B_{\rho\mu}
\ ,\ \ \ F_{\mu\nu}=\partial_\mu A_\nu-\partial_\nu A_\mu \nonumber
\end{equation}
Here  $M$
is a mass scale (of order the string scale $M_s$  in string models) and  c
is a model dependent constant of order one.
  The last term in this expression
is the $B\wedge F$ coupling mentioned  above.  Let us now
review how the gauge boson gets massive by combining with the
axion-like field $\eta$ (see e.g. \cite{giiq}).  
We can rewrite this Lagrangian in terms of
$H_{\mu \nu \rho}$ imposing the constraint $H=dB$ introducing a
Lagrange Multiplier $\eta $:
\begin{equation}
{\cal L}_0\ =\ -\frac{1}{12} H^{\mu\nu\rho}
H_{\mu\nu\rho}-\frac{1}{4g^2} F^{\mu\nu} F_{\mu\nu}
- \frac{cM}{6}\ \epsilon^{\mu\nu\rho\sigma} H_{\mu\nu\rho}\ A_\sigma
-\frac{cM}{6}\eta \epsilon^{\mu\nu\rho\sigma}\partial_\mu H_{\nu\rho\sigma}
\label{lagranb}
\end{equation}
Now we can use the equations of motion for $H_{\mu\nu\rho}$
and find
\begin{equation}
H^{\mu\nu\rho}\ =\ -cM \epsilon^{\mu\nu\rho\sigma}(A_\sigma +\partial_\sigma
\eta ) 
\label{hache}
\end{equation}
Substituting back into eq.(\ref{lagranb}) one obtains
\begin{equation}
{\cal L}_M\ =\ -\frac{1}{4g^2} F^{\mu\nu} F_{\mu\nu}
- \frac{c^2M^2}{2}(A_\sigma +\partial_\sigma \eta )^2
\label{lagranc}
\end{equation}
This corresponds to the Stuckelberg Lagrangian of a massive vector boson
of mass $cgM$.

Let us now proceed to the consideration of $H_{\mu \nu \rho }$ fluxes in a
scheme with a gauged  $U(1)$ as above. In principle such a vev
would explicitly violate Lorentz invariance. However we will   
turn on fluxes only along the three space dimensions so that
 at the cosmological level there will be no contradiction
with experimental facts
\footnote{Cosmology  with a  flux 
$H_{\mu\nu\rho}$ along the three space dimensions
has been studied  in the past, see e.g. \cite{lwc} and references therein.   
However no connection with a baryon asymmetry was considered.}.
We will assume now a non-vanishing constant value of $H$,
$h_{\mu\nu\rho}$. Then eq.(\ref{lagranb}) is modified to
\begin{equation}
{\cal L}\ =\  {\cal L}_0 \ -\ \frac{1}{12} h^{\mu\nu\rho}
h_{\mu\nu\rho}-
 \frac{cM}{6}\ \epsilon^{\mu\nu\rho\sigma} h_{\mu\nu\rho}\ A_\sigma
\label{lagrand}
\end{equation} 
and  the final Lagrangian has the form
\begin{equation}
{\cal L}\ =\  {\cal L}_M \ -\ \frac{1}{12} h^{\mu\nu\rho}
h_{\mu\nu\rho}-
  J_H^\sigma \ A_\sigma
\ -\ J_F^\sigma \ A_\sigma
\label{lagrane} 
\end{equation}
where  
\begin{equation}
J_H^\sigma \ =\ \frac{cM}{6} \epsilon^{\mu\nu\rho\sigma} h_{\mu\nu\rho}
\label{corrhache}
\end{equation}
and we have added a term corresponding to the current $J_F^\sigma$ 
of the fermions coupling to the $U(1)$ gauge boson.
In summary, we get a massive gauge boson but in addition the 
flux background $h_{\mu\nu\rho}$ acts like a current coupling to
the massive $U(1)$ gauge boson. We will assume that the $H$-background 
is present in the universe only
for the three spacelike components x,y,z of the
flux
\begin{equation}
 h_{xyz}\ = \ H\epsilon _{xyz}\not= \ 0
\end{equation}
so that actually the flux  induces a  $U(1)$-charge density.
Although the $U(1)$
gauge boson is  massive, we already pointed out that  an
unbroken {\it global} $U(1)$  symmetry  persists.
 Thus, at the level
of the low-energy effective Lagrangian (below the scale of
the gauge  boson mass) the effect of a vev for $H$
is to induce a non-vanishing {\it global} $U(1)$
 charge density.

The above discussion applies to any gauged $U(1)$
symmetry whose gauge boson becomes massive a
la Stuckelberg. This $U(1)$ may be anomaly free 
(like B-L) or anomalous, with the anomaly being
cancelled by the Green-Schwarz mechanism.
We want to apply   these ideas to the case of the
baryon number of the universe. Let us consider then for 
simplicity the case of a 
gauged   $U(1)_{B-L}$ symmetry.
In this situation we will have in the early universe
a net vanishing  B-L  charge. Below the scale
at which the $U(1)_{B-L}$ gauge boson gets a mass
(i.e. the string scale in string models) (B-L)- number
survives  as a conserved global symmetry. Since  this symmetry was gauged,
in the primordial universe the overall B-L  charge 
should vanish, very much like electric charge should
vanish. Then the conservation of the residual global $U(1)_{B-L}$
current dictates that at the level of the
effective field theory one has for the matter 
B-L  density $n_{B-L}$
\begin{equation}
   n_{B-L}\ +\ cMH  \ =\ 0
\label{conscor}
\end{equation}
Thus as long as we have $H\not= 0$ there will be a
a residual B-L  number in the form of baryons/leptons  
and the `B-L aether' induced by $H$ will
compensate to get  an overall  vanishing (B-L) number.
Note that both charges will be conserved separately.
On the other hand at some point, at lower temperatures 
the electroweak instanton effects may  be in
thermal equilibrium giving rise to (B+L) violation
in the standard way. The only effect of these will be to 
make $n_B=-n_L$ and the baryon-antibaryon 
asymmetry will persist.

One interesting question is what is the contribution of 
a non-vanishing $H$ to the present energy density in the universe.
Is it sufficiently big to account for the observed 
cosmological constant? The answer is no, it contributes
in a negligible way to the present energy density.
Since $H$ and baryon densities are related
by $|H| \simeq  |n_B| /M$
one can easily make an estimate.
As seen in eq.(\ref{lagrand}) one expects a contribution to the vacuum energy
$V_H\propto H^2$ and hence one expects for the ratio
of densities from $H$ and from baryons
\beq
\frac {\Omega_H}{\Omega_B} \ \propto
\frac {H^2}{\rho_B} \ =\ \frac {\rho_B}{ ( m_p^2 M^2)}
\eeq
At present temperatures this ratio is extremely small for any
reasonable value of the fundamental scale $M$ (
one has ${\Omega_H}/{\Omega_B}\propto
10^{-80}$ for $M= 10^{16}$ GeV). Thus the contribution of the  flux
$H$ to the present vacuum energy density seems negligibly small.
However, this contribution to the vacuum energy
may have been much more important in the past.
The reason for this is that \cite{Nastase} a background for
$H$ in the Einstein's equations behaves like `stiff matter'
(i.e., $p=\rho$)
 so that  one has $\rho _H\propto 1/a(t)^6$,
$a(t)$ being the scale factor. 
Compared to the baryon density one thus have
\beq
\frac {\rho_H}{\rho_B} \ \propto \ \frac {1}{a(t)^3} \ .
\eeq
Note that this  behavior is consistent with equation
(\ref{conscor}) since it implies
that $H^2$ scales with the scale factor like $n_B^2\propto
\frac {1}{a(t)^6}$, as expected.
Thus the evolution equations are consistent with
eq.(\ref{conscor}) and the conservation of baryon number at
any time.

The flux vev $H$ is in principle a free parameter of the underlying
theory, very much like other fluxes considered recently in the
context of string theory in order to stabilize the moduli
\cite{grana}.
An important difference is that these other fluxes go through
the compactified extra dimensions whereas the flux here considered
goes through the three space dimensions and has direct cosmological
relevance. It would be interesting if we could figure
out what is a natural value for the density $H$  since, 
given eq.(\ref{conscor}),  we could then compute the baryon asymmetry.
It could well be that the density  $H$ could be determined 
on anthropic grounds, certainly our existence very much
depends on the amount of baryonic matter.
On the other hand it would
be interesting to have 
a  specific model of string inflation in which the correct size 
for $H$ was dynamically determined.
The H-background had to appear after inflation, otherwise
it would have been  totally diluted. If it was created say at the 
reheating temperature $T^*$,  one can estimate the 
baryon asymmetry density then to be
\beq
\frac {n_B}{n_\gamma}\ \simeq \
\frac {M\ H}{(T^*)^3}
\label{asi1}
\eeq
If we insist in getting an asymmetry $n_B/n_\gamma
\simeq 10^{-10}$ one would need to have a flux
\beq
H \ \simeq \ \frac{(T^*)^3}{M} 10^{-10}
\label{hflation}
\eeq
If  $M$ of order $10^{16}$ GeV (corresponding to a string scale
of order the GUT scale) and a reheating temperature say of
order $10^9$ GeV  then the required flux at reheating 
is of order of a hadronic scale, $H\simeq (300\ MeV)^2$.
If $M$ is of order the intermediate scale 
$M\simeq 10^{11}$ GeV (as advocated 
in some string models), then the required flux 
is of order  $H\simeq (1\ Tev)^2$. 
One can play around with different values for the 
reheating temperature and the string scale leading to
the desired asymmetry.

If one could raise up the reheating scale close to 
a string scale of order $10^{16}$  GeV  one could
relate the  asymmetry to the scale of SUSY-breaking.
Although this sounds unlikely, let us explain it for the 
sake of the argument. Consider 
 the context of type II orientifold string compactifications,
which is a natural setting for the present mechanism.
As we said, appart from this flux,  in generic compactifications
there are other antisymmetric fluxes  
(let me call them collectively  $G$)
which wrap cycles in the 
compact dimensions \cite{grana}.
 This is   a crucial ingredient in recent
efforts 
in order to understand the dynamical fixing of the string
moduli. In addition to fixing the moduli, it has also been shown 
\cite{soft} that
generically such fluxes $G$ do also break supersymmetry and give rise
to SUSY-breaking soft terms of order $G/M_p$. In order to obtain
soft terms of order $M_{sb}\propto 1$ TeV  the fluxes must be diluted and
be $G\propto M_{sb}M_p$. On the other hand at temperatures close to  
$T^*\simeq M$ it is natural to expect that both these fluxes
and the one considered in this paper have similar densities,
$H\simeq  G\simeq  M_{sb}M_p$, since at that scale
there is not much difference between compact and non-compact 
fluxes. If this is the case we would 
obtain for the asymmetry
\beq
\frac {n_B }{n_\gamma}\ \simeq \
\frac {H}{M^2} \ \simeq \ \frac {M_{sb}M_p}{M^2}
\label{asi2}
\eeq
For SUSY-breaking soft masses of $M_{sb}\propto 1$ TeV and
  $M$ of order $10^{16}$ GeV (corresponding to a string scale
of order the GUT scale) this ratio comes to be of order
$10^{-10}$, of the order of the  observed 
asymmetry. 
Note that if dark matter is constituted by SUSY neutralinos,
both baryon asymmetry and dark matter would then be correlated since
both would depend sensitively on the scale of SUSY-breaking. 
Let us however emphasize that this numerical exercise
with such very high reheating temperature sounds
unlikely within the present known models for reheating.

In this note we have emphasized the case of a gauged B-L symmetry
because of its simplicity, it is anomaly-free and requires no
Green-Schwarz mechanism. 
But is clear that the mechanism generalizes to 
the gauging of other possible global symmetries of the SM
or any of its extensions,
e.g. other  linear combinations of B and L. The necessary ingredients
are 1) a gauged $U(1)$ symmetry (anomalous or not) whose
gauge boson gets a mass term a la Stuckelberg and 2) a nonvanishing
flux across the space dimensions for the corresponding 
antisymmetric field. Under those conditions the $U(1)$ symmetry
survives as a global symmetry and the flux induces an asymmetry of the 
corresponding charge. The only constraint is that the linear combination 
should be different from (B+L) because in this case electroweak
instantons would erase any primordial asymmetry.
One can also consider some other $U(1)$'s coming from some hidden sector
of the theory, not coupling directly to the SM fields. In this case this
mechanism could give rise to some density of hidden sector particles
which could play the role of dark matter. If the relevant flux is 
of the same order of magnitude than that  generating baryons and the masses 
of those dark matter objects is one order of magnitude larger than that
of baryons, this could explain why dark and visible matter turn out to have
not very different contributions to the energy of the universe.

\vspace{2.0cm}

\centerline{\bf Acknowledgements}

\vspace{0.5cm}

I  thank P.G. C\'amara, D. Cremades, J. Garc\'{\i}a-Bellido,  J. Garriga,
A. Font, F. Marchesano, F. Quevedo, R. Rabad\'an and A. Uranga for useful
comments and discussions. This work has been partially supported by CICYT
(Spain)and the European Commission under the RTN program
MRTN-CT-2004-503369.

\vspace{2.0cm}


\end{document}